\documentclass[conference]{IEEEtran}
\IEEEoverridecommandlockouts
\pdfoutput=1
\usepackage{cite}
\usepackage{amsmath,amssymb,amsfonts}
\usepackage{algorithmic}
\usepackage{graphicx}
\usepackage{textcomp}
\usepackage{diagbox}
\usepackage{multirow}
\usepackage{balance}
\usepackage{xcolor}

\usepackage{hyperref}
\usepackage{tabularx} 
\usepackage{booktabs} % For formal tables
\usepackage{hhline}
\usepackage{array}
\usepackage{xcolor}
\usepackage{longtable}
\usepackage{float}
\usepackage{svg}
\usepackage{enumitem}
\usepackage{scrextend}

\newcommand{\revision}[1]{#1}
\newcommand{\code}[1]{{\tt #1}}

\newcolumntype{L}[1]{>{\raggedright\let\newline\\\arraybackslash\hspace{0pt}}m{#1}}
\newcolumntype{C}[1]{>{\centering\let\newline\\\arraybackslash\hspace{0pt}}m{#1}}
\newcolumntype{R}[1]{>{\raggedleft\let\newline\\\arraybackslash\hspace{0pt}}m{#1}}

\def\BibTeX{{\rm B\kern-.05em{\sc i\kern-.025em b}\kern-.08em
    T\kern-.1667em\lower.7ex\hbox{E}\kern-.125emX}}

\newenvironment{boxedtext}
    {
    
    \begin{center}

    \begin{tabular}{|p{0.96\linewidth}|}
    \hline
    }
    { 
    \\ \hline
    \end{tabular} 
    
    \end{center}
       }

\begin{document}

\title{Towards Automated Classification of Code Review Feedback to Support Analytics}

\author{\IEEEauthorblockN{Asif Kamal Turzo $^\heartsuit$,
    Fahim Faysal $^\clubsuit$,
        Ovi Poddar  $^\clubsuit$
        Jaydeb Sarker $^\heartsuit$\\
        Anindya Iqbal  $^\clubsuit$
        Amiangshu Bosu $^\heartsuit$}
\IEEEauthorblockA{
$^\heartsuit$\textit{Wayne State University, Detroit, Michigan, USA}\\
\textit{$^\clubsuit$Bangladesh University of Engineering and Technology, Dhaka, Bangladesh}\\
\textit{asifkamal@wayne.edu},
\textit{ffalsoy@gmail.com},
\textit{ovi.poddar2@gmail.com},
\textit{jaydebsarker@wayne.edu},\\ 
\textit{anindya@cse.buet.ac.bd},
\emph{amiangshu.bosu@wayne.edu}} 
}

%\IEEEoverridecommandlockouts
%\IEEEpubid{\makebox[\columnwidth]{978-1-6654-5223-6/23/\$31.00 ©2023 IEEE \hfill}
%\hspace{\columnsep}\makebox[\columnwidth]{ }}

\maketitle

\begin{abstract}
\revision{\textit{Background:} As improving code review (CR) effectiveness is a priority for many software development organizations, projects have deployed CR analytics platforms to identify potential improvement areas. The number of issues identified, which is a crucial metric to measure CR effectiveness, can be misleading if all issues are placed in the same bin. Therefore, a finer-grained classification of issues identified during CRs can provide actionable insights to improve CR effectiveness. Although a recent work by Fregnan \emph{et} al. proposed automated models to classify CR-induced changes, we have noticed two potential improvement areas -- i) classifying comments that do not induce changes and ii) using deep neural networks (DNN)  in conjunction with code context to improve performances.}

\revision{\textit{Aims:} } This study aims to develop an automated CR comment classifier that leverages DNN models to achieve a more reliable performance than Fregnan \emph{et} al.

\revision{\textit{Method:} } Using a manually labeled dataset of 1,828 CR comments, we trained and evaluated supervised learning-based DNN models leveraging code context, comment text, and a set of code metrics to classify CR comments into one of the five high-level categories proposed by Turzo and Bosu. 

\revision{\textit{Results:} } \revision{Based on our 10-fold cross-validation-based evaluations of multiple combinations of tokenization approaches, we found a model using CodeBERT achieving the best accuracy of 59.3\%. Our approach outperforms Fregnan \emph{et} al.'s approach by achieving 18.7\% higher accuracy.}

\revision{\textit{Conclusion:} } \revision{In addition to facilitating improved CR analytics, our proposed model can be useful for developers in prioritizing code review feedback and selecting reviewers.}
\end{abstract}


\section{Introduction}
\label{sec:intro}
\revision{Peer Code Review (CR) is ubiquitous among contemporary software development pipelines. CR acts as a quality assurance gateway among popular Open Source Software (OSS) and commercial organizations such as Microsoft, Google, and Facebook~\cite{rigby2014peer,bosu2016process}~\cite{sadowski2018modern,rigby2014peer}. 
Since many projects have made CR mandatory~\cite{rigby2013convergent,wurzel2023competencies}, developers spend, on average, 10-15\% of their time (i.e., more than an hour each day) on CR tasks~\cite{bosu2016process}. While the primary expectation behind CR adoption is finding defects, most CRs do not meet this expectation~\cite{bacchelli2013expectations}. Recent studies found four out of the five CR comments regarding style and nitpicking issues~\cite{bosu2015characteristics,turzo2023makes,beller2014modern} that can also be identified using static analysis tools. Therefore, some project managers often wonder if their CRs are cost-effective and worth practicing~\cite{czerwonka2015code,bacchelli2013expectations}.On the other hand, some project managers, although they consider CRs crucial due to other benefits such as knowledge dissemination~\cite{sadowski2018modern,bacchelli2013expectations}, they seek to improve CR effectiveness, as even a minor improvement can incur significant savings for  large organizations such as Microsoft and Google~\cite{bosu2015characteristics,sadowski2018modern}}.

\revision{To identify potential improvement areas, many organizations have developed dashboards detailing various CR analytics such as review completion time, best reviewers, and the number of issues identified~\cite{hasan2021using,bird2015lessons}. Although the number of issues identified during CRs is an important metric to measure CR effectiveness, this metric, as measured by existing analytics tools~\cite{hasan2021using}, can be misleading due to the coarse-grained analysis of identified issues. For example, a CR comment identifying a critical functional defect is placed in the same bin as one suggesting nitpicking issues such as fixing typos. 
Therefore, the ratio of functional defects as opposed to nitpicking ones among the identified issues can not be measured. If CRs mostly identify nitpicking issues, yet bugs are identified post-CR, there may be shortcomings in the current CR process that current CR analytics would fail to reveal. An automated classifier is necessary to facilitate more informative CR analytics as manual classification of CR comments on a large scale is infeasible. }

On this need, Fregnan \textit{et} al.~\cite{fregnan2022happens} developed an automated model to classify CR-induced changes into four categories. The results of their user evaluation show a need for such a CR analytics platform among the developers. \revision{However, we noticed two potential improvement areas over their work. First, as their model focuses on classifying review-induced changes, it fails to account for CR comments that do not trigger changes (e.g., discussion or changes deferred as future works), even though that feedback was beneficial. Therefore, analytics based on Fregnan \textit{et} al.'s \cite{fregnan2022happens} model fail to allocate credits to authors of such reviews. Second, Fregnan \textit{et} al.~\cite{fregnan2022happens} uses only classical Machine Learning (ML) algorithms such as Naive Bayes, Decision Tree, and Random Forest in their evaluation. We aim to explore deep neural network (DNN)-based algorithms with CodeBert~\cite{feng2020codebert}, as these combinations have shown superior performances for code classification tasks~\cite{zhou2021assessing,liu2022codebert}.}  Therefore, this study aims to develop an automated CR comment classifier that leverages DNN models to achieve a more reliable performance than Fregnan \emph{et} al.~\cite{fregnan2022happens}

On this goal, we have manually labeled a dataset of 1,828 CR comments following the CR comment classification scheme proposed by Turzo and Bosu~\cite{turzo2023makes}. Using this dataset, we trained and evaluated supervised-learning-based DNN models leveraging various code and feedback characteristics to classify each CR comment into one of the five high-level categories proposed by Turzo and Bosu~\cite{turzo2023makes}. We empirically evaluated multiple combinations of tokenization approaches to identify the best-performing model. In summary, we answer the following two research questions:

\noindent \textbf{(RQ1) Can we use an automated machine learning-based approach to classifying code review comments?}

\noindent \underline{Motivation:}  \revision{An automatic CR comment classification model would provide three benefits. First, an author can prioritize high-priority issues to prepare revisions. Second, based on prior CR analytics, an author may identify reviewers who are more likely to identify their priority issues and select reviewers accordingly. Finally, it can provide project managers with additional CR analytics to access individuals and overall process performances.}
 
 \vspace{2pt}
  \noindent \underline{Method:} We develop a machine learning-based classification approach that considers code context, CR comments, and a set of code attributes for classifying CR comments.
  
 \vspace{2pt} 
 \noindent \underline{Results:} The proposed approach can classify CR comments with an overall accuracy of $59.3\%$. Our evaluation also suggests that all the features (i.e., code context, comment text, and code attributes) improve model performance.

\noindent \textbf{(RQ2) Can our review comment classification tool perform better than the current state-of-the-art?}

\noindent \underline{Motivation:} \revision{Review comment classification task is slightly different from change classification because some comments might not induce change. However, by retraining, the existing change classification approach of Fregnan et al.\cite{fregnan2022happens} can be used for classifying review comments. They considered code attributes only for classifying a change. Our hypothesis is \textit{if we incorporate code context, review comment, and additional code attributes, we may achieve better performance to classify review comments}. RQ2 tests this hypothesis.}
 
 \vspace{2pt}
  \noindent \underline{Method:} We replicated the study of Fregnan \emph{et} al.\cite{fregnan2022happens} on our dataset of 1,828 CR comments from the OpenDev Nova project. We used precision, recall, F1-score, and model accuracy to compare the performance of both approaches.
  
 \vspace{2pt} 
 \noindent \underline{Results:} Experimental results found that our proposed approach achieved an 18.7\%  performance improvement in model accuracy  over Fregnan \emph{et} al.'s approach in the review comment classification task.

 \vspace{4pt}
 The primary contributions of this study are the following:
\begin{itemize}
  \item A novel machine learning approach for classifying CR comments.
  \item A detailed investigation by varying the input attributes and experimenting with the effect of different review comment tokenization and vectorization approaches on the model's performance.
  \item Empirically validated that the proposed approach performs better than the existing comparable automatic change classification approach of Fregnan \emph{et} al.\cite{fregnan2022happens}.
  \item We have made our code and dataset publicly available at: \url{https://github.com/WSU-SEAL/CR-classification-ESEM23}
\end{itemize}

%Paper outline
The remainder of the paper is organized as the following. Section~\ref{sec:background} discusses prior \revision{related} works. 
Section \ref{sec:dataset} presents the data collection and manual labeling approach. Section \ref{sec:rq1} details our model construction and evaluation approach and answers RQ1. 
Section \ref{sec:rq2} answers RQ2 by comparing the performance of our approach with the approach of Fregnan et al.\cite{fregnan2022happens}. 
Section \ref{sec:discussion} discusses the insight gathered from this experiment. Section \ref{sec:thrests} and \ref{sec:conclusion} address the threats to validity and concludes the paper, respectively.

\section{Related Works}
\label{sec:background}
The following subsections briefly describe prior research relevant to the goals of this study.
\vspace{2pt}

\noindent \revision{\textbf{Code review effectiveness :}} Due to the widespread adoption of code review, recent years have seen increased research focus on this area. This mismatch between developers' expectations from CRs and outcomes is due to code comprehension challenges with limited time and context~\cite{bacchelli2013expectations}. Their finding that approximately 80\% CRs do not find bugs is also supported by several subsequent studies~\cite{bosu2015characteristics,czerwonka2015code,hasan2021using,beller2014modern,turzo2023makes}.
Although most CRs do not find bugs, developers still consider CR as an essential practice due to other benefits such as knowledge dissemination, improving project maintainability, and community building~\cite{kononenko2016code,bosu2016process,sadowski2018modern}.
Studies investigating CR effectiveness have identified several technical and non-technical factors having positive associations, which include the reviewer's experience~\cite{turzo2023makes,bosu2015characteristics,thongtanunam2016revisiting}, the author's reputation~\cite{bosu2014impact}, and project tenure~\cite{bosu2015characteristics}. On the other hand, factors with negative associations with CR effectiveness include missing rationale, discussion of non-functional requirements of the solution, and lack of familiarity with existing code~\cite{ebert2021exploratory}, co-working frequency of a reviewer with the patch author ~\cite{thongtanunam2020review,turzo2023makes}, description length of a patch~\cite{thongtanunam2017review}, changeset size~\cite{rigby2014peer}, the number of files under review~\cite{bosu2015characteristics,turzo2023makes} and the level of agreement among the reviewers~\cite{hirao2016impact}.

\vspace{4pt} \noindent \revision{\textbf{Code review automation:}}  To help the reviewers understand the code better, several studies have focused on developing the changeset decomposition tool \cite{barnett2015helping,dias2015untangling,gomez2015visually,huang2020code}. By decomposing the changeset, these studies aim at helping reviewers understand the code better so that reviewers can provide useful functional feedback. A large number of prior studies have developed tools for selecting appropriate reviewers \cite{zanjani2015automatically,thongtanunam2015should,rahman2016correct,pandya2022corms,rong2022modeling}. By finding appropriate reviewers, those tools assist code authors in obtaining useful review comments. Existing reviewer recommendation systems consider history as a benchmark for measuring performance. But a recent study suggests that history can overestimate or underestimate the capability of a reviewer recommendation system \cite{gauthier2021historical}. Several recent studies have proposed the complete automation of the whole code review process \cite{thongtanunam2022autotransform,tufano2021towards,li2022codereviewer,tufano2022using,tufano2019learning}. With the recent advancement of transformer model\cite{vaswani2017attention}, automation of code review activity can be achieved. But, such complete automation of code review activity would deprive developers of the benefits of code review, such as knowledge sharing among team members. \revision{We are motivated by prior studies that found the need to develop tools for automating different aspects of the code review process and developers' interests in CR analytics platforms~\cite{bird2015lessons,hasan2021using,fregnan2022happens}.}

\vspace{4pt} \noindent \revision{\textbf{Code review classification:}} Several studies have manually classified CR comments to identify the ratios of various categories of identified issues. Mäntylä and Lassenius first studied industrial and student CRs and found that only one-fourth of CRs affect code functionality \cite{mantyla2008types}. They found 12 types of  CR-identified issues, which they further categorized into three higher-level groups. Beller \emph{et} al. manually classified 1,400 CR changes and found 10-22\% of changes unrelated to CR comments \cite{beller2014modern}. They also proposed a taxonomy that categorized CR-induced changes into two higher-level categories. As some of the CR comments do not induce any change, prior studies also focused on classifying CR comments. Bosu \emph{et} al. conducted a study on Microsoft where they identified 13 types of comments that appear in the code review process \cite{bosu2015characteristics}. Turzo and Bosu \cite{turzo2023makes} recently adopted Beller \emph{et} al.'s change classification scheme to create a 
 CR comment taxonomy consists  of 17 subcategories grouped into  five higher-level categories. 
Using a labeled dataset of 800 CR comments, Bosu \emph{et} el.~\cite{bosu2015characteristics} trained an automated classifier to automatically identify useful CR comments. Building on their work, two later studies have developed classifiers with the same goal at different  settings~\cite{rahman2017predicting,hasan2021using}.
 While earlier studies focused on a binary classification to predict whether a CR comment was useful, Fregnan \emph{et} al.~\cite{fregnan2022happens} is the first to propose an automated model to classify review-induced changes into four categories. As Fregnan \emph{et} al.'s~\cite{fregnan2022happens} model focuses on classifying review-induced changes, our work also differs from theirs by considering non-change inducing CR comments.
\section{Dataset Preparation}
\label{sec:dataset}

\begin{figure}
	\centering  \includegraphics[width=\linewidth]{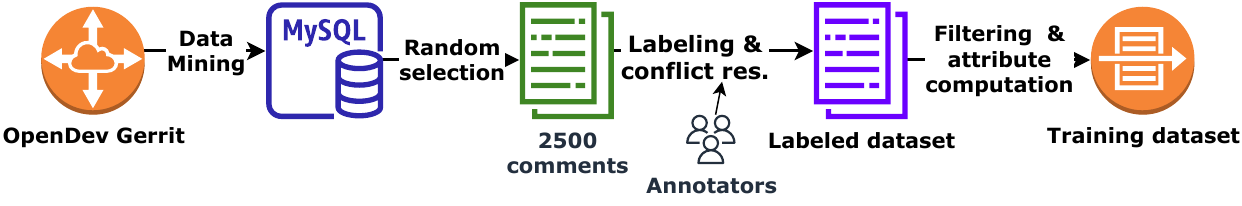}
\caption{An overview of our dataset preparations steps}
\label{fig:data-prepare}
\vspace{-14pt}
\end{figure}

This section details our project selection, data mining, CR comment classification rubric, and manual labeling approach. Figure~\ref{fig:data-prepare} shows an overview of our dataset preparation steps.

\subsection{Project Selection}
For this study, we selected the OpenDev Nova project from the OpenDev community for the following reasons:

\begin{enumerate}
    \item Developers in the OpenDev community practice tool-based CR \cite{bosu2016process}, and OpenDev is one of the largest  OSS communities with contributors from more than 700 organizations~\footnote{https://openinfra.dev/annual-report/2022} worldwide. 

\item OpenDev projects have been the subject of prior CR studies \cite{zanaty2018empirical,turzo2023makes,hirao2019review}.
\end{enumerate}

\subsection{Dataset Preparation}
OpenDev community uses Gerrit\footnote{https://www.gerritcodereview.com/} to manage CRs. We accessed the REST API provided by Gerrit to access and mine all the publicly available CRs from OpenDev's Gerrit instance\footnote{https://review.opendev.org}.
Our dataset spans July 2011 to March 2022 and includes a total of 795,226 either `Merged' or `Abandoned' CRs. We stored the dataset in a local MySQL server. 

\begin{table*}
\caption{Rubric for classifying code review comments which we adopted from prior studies \cite{beller2014modern,mantyla2008types,bosu2015characteristics,turzo2023makes,hasan2021using}}
\centering \label{table:rubric}
\resizebox{\linewidth}{!}{  
\begin{tabular}{ |p{2cm} |p{2.5cm} | p{12.1cm}| } 
\hline
\textbf{Group} & \textbf{Category} & \textbf{Description}   \\ 
\hline
\multirow{7}{*}{\textbf{Functional}}
& Functional  & Functional issues are defects where a code functionality is missing or implemented incorrectly. If requires large modification to resolve the issue. \\ \cline{2-3}
& Logical  & Logical issues are defects where there exist control flow problems or logical mistakes (wrong logic implementation, comparison issues). \\ \cline{2-3}
& Validation  & All types of user data sanitization issues or issues related to exception handling.\\ \cline{2-3}
& Resource  & Any kind of variable, memory, or file issues while handling or manipulating them. \\ \cline{2-3}
& Timing & Any kind of synchronization issues while using the thread. \\ \cline{2-3}
& Support issues & Any kind of support systems-related issues (e.g. configuration problem or version mismatch).  \\ \cline{2-3}
& Interface  & Any types of interfacing issues such as issues in an import statement, issues while interacting with the database or internal system.
 \\ \hline
\multirow{5}{*}{\textbf{Refactoring}}
& Solution approach & Review comments that suggest an alternative approach for problem-solving. \\ \cline{2-3}
& Alternate Output  & Review comments that address issues within the alert message, toast message, or error message. \\ \cline{2-3}
& Code Organization  & Code organization or refactoring issues presented in the catalog of Martin Fowler\cite{fowler2012refactoring}. \\ \cline{2-3}
& Variable Naming  & Review comments that address the violation of the variable naming convention. \\ \cline{2-3}
& Visual Representation & Any kind of indentation, blank line, or code spacing-related issues. \\  \hline
 \textbf{Documentation} & 
 Documentation  & Review comments that address issues related to code comments or documentation files for aiding code comprehension. \\ 
\hline

\multirow{3}{*}{\textbf{Discussion}} & Design discussion & Comments that discuss design pattern or sourcecode architecture. \\ \cline{2-3} 

 & Question & If reviewers ask anything to the code author for clarification. \\ \cline{2-3}
 & Praise & Review comments that praise or complement the developer.\\ \hline

\textbf{False positive} & False positive & An invalid concern. \revision{A CR comment is considered `False Positive' if (a) the code owner explicitly mentions the comment as an invalid concern or (b) no subsequent change occurred nor the code owner agrees to a future change \cite{turzo2023makes}.} \\ 
\hline

%\textbf{Others}  & Other & Comments not belonging to any of the above categories.\\
%\hline

\end{tabular}
}
\vspace{-16pt}
\end{table*}

\subsection{Manual Labeling}
We randomly selected 2,500 CR comments from the OpenDev Nova project for manual labeling using MySQL's {\tt rand()} function. \revision{We only selected CR comments from OpenDev Nova as it has the highest number of posted CRs during our dataset span of  July 2011 to March 2022}. 
Prior studies have proposed several CR issues classification schemes~\cite{mantyla2008types,beller2014modern,bosu2015characteristics,turzo2023makes,hasan2021using,fregnan2022happens}. In this study, we select the classification scheme proposed by Turzo and Bosu~\cite{turzo2023makes}, as their classification considers CR comments that did not induce any change, opposed to the one used by Fregnan \emph{et} al.~\cite{fregnan2022happens} Since we aim to classify CR comments, Turzo and Bosu's scheme is more suitable for our goal. Table ~\ref{table:rubric} shows the 17 CR comment subcategories that are grouped into five higher-level categories based on this classification scheme.
As building a reliable 17-category classifier is challenging, this study focuses on the five higher-level categories, which  are: i) \code{Functional},  \code{Refactoring}, iii) \code{Documentation'} iv) \code{Discussion}, and v) \code{False Positive}.

\revision{Although our classifier requires labeling each CR comment into one of the five high-level categories, we manually labeled using the 17-category classification for future use cases of this dataset and fine-grained error analysis.} Each of the CR comments was independently labeled by two annotators. During this process, they read the entire comment thread and its surrounding code context and induced changes among subsequent reviews (if any) to assign one of the  17 labels (Table~\ref{table:rubric}). We compared the assigned labels to identify conflicts. We computed  Cohen's kappa ($\kappa$)~\cite{cohen1960coefficient} to measure the inter-rater reliability of this manual labeling process. The manual labeling achieved a $\kappa$ value of 0.68 which indicates a substantial agreement\footnote{Kappa ($\kappa$) value interpretation: 0.01–0.20 indicates `none to slight', 0.21–0.40 indicates `fair', 0.41– 0.60 indicates `moderate', 0.61–0.80 indicates `substantial', and 0.81–1.00 indicates `almost perfect agreement'~\cite{landis1977application}.}. 
To resolve the conflicts, a third annotator independently went through the conflicting ones to assign final labels. \revision{All three annotators are co-authors of this paper and have significant research experience in code reviews.} After completing the manual labeling step, we assign the target label for each CR comment according to our schema from Table~\ref{table:rubric} . For example, if a category comment belongs to `Design discussion', `Question', or `Praise', we assign the comment to the \code{Discussion} group. We found that 672 among the 2,500 CR comments were unrelated to source code files (e.g., configuration, build, resources, and commit message). As many of the code attributes selected for our classifiers (Table~\ref{table:features}) cannot be computed for non-source code files, we exclude those 672 CR comments at this step. After this exclusion, we use the remaining 1,828 CR comments to train and evaluate our classifiers. \revision{In this dataset,  8.64\% comments belong to \code{False Positive}, 24.34\% belong to \code{Discussion}, 32.71\% belongs to \code{Refactoring}, 21.17\% belongs to \code{Documentation}, and 13.13\% comments belong to \code{Functional} group.}

\section{(RQ1) Machine Learning Model for classifying code review comments}
\label{sec:rq1}

 To classify CR comments, we consider three types of input attributes for our machine-learning model. We consider the code context where the review comment was made, the CR comment in the form of natural language, and several code attributes. This section presents our attribute selection, model training approach, and results of our evaluations.

\begin{table*}
\caption{Code attributes selection for machine learning modeling}
\centering \label{table:features}
\resizebox{\linewidth}{!}{
\begin{tabular}{p{2.9cm} p{5.5cm} p{7.5cm} }
\hline
\textbf{Attributes}            & \textbf{Description} & \textbf{Rationale} \\ \hline
\multicolumn{3}{l}{\textbf{AST-based attributes}}\\ \hline
anyInserted  & Count of AST nodes inserted in the \emph{destination} file. & Some code changes involve the addition of a few AST nodes. For example, in case of variable value assignment or logical changes. \\ \hline
anyDeleted & Count of AST nodes deleted in the \emph{destination} file.  & Some code changes involve the deletion of a few AST nodes. In case of variable value assignment or logical changes, few AST node deletions might occur. \\ \hline
getMovedSrcs & How many AST nodes moved from the \emph{source} file.  & If the \emph{source} code chunk is moved to the \emph{destination}, it might indicate a refactoring change. \\ \hline
updatedSrcs & How many \emph{sources} AST nodes are updated in the \emph{destination} file?  & Update of a few numbers of AST nodes may indicate logical changes. \\ \hline
anythingInLineMoved  & Count of \emph{source} AST nodes that were within the Review Comment Range (RCR) but moved elsewhere in the \emph{destination} file. & Prior studies suggest most changes occur in close proximity to the review comment. The number of AST nodes moved within close proximity can provide valuable information about the change. \\ \hline
anythingInLineUpdated & Count of \emph{source} AST nodes that were within the RCR and were updated in the \emph{destination} file. & Similarly, the number of AST nodes updated within close proximity of the review comment can provide valuable information about the change. \\ \hline
anythingInLineDeleted & Count of \emph{source} AST nodes that were within the RCR and were deleted.  & Similarly, we argue that the number of AST nodes deleted within close proximity of the review comment can provide valuable information about the change.  \\ \hline
anythingMovedIntoLine & If an AST node is in the RCR of the \emph{source}, then the number of child nodes moved within that node in the \emph{destination} file.  &  Number of nodes moved within RCR can provide crucial information for review comment classification, it can be indicative of refactoring changes. \\ \hline
anythingInsertedIntoLine &  If an AST node is in the RCR of the \emph{source}, then the number of child nodes that are newly inserted within that node in the \emph{destination} file. & Number of nodes newly inserted within RCR can provide crucial information for review comment classification, it can be indicative of larger changes. \\ \hline
\multicolumn{3}{l}{\textbf{Change-based attributes}}\\ \hline
insertedIfConditions & How many if statements in the \emph{destination} file were inserted?  & Insertion of `if' condition indicates logical Functional change.\\ \hline
deletedIfConditions & How many if statements in the \emph{source} file were deleted? & Similarly, deletion of the `if' condition indicates logical Functional change. \\ \hline
elseInserted & How many else inserted in the \emph{destination} file? & Insertion of `else' condition indicates control flow change, which is a Functional change. \\ \hline
elseDeleted & How many else statements in the \emph{source} file were deleted?  & Similarly, deletion of the `else' condition indicates logical Functional change. \\ \hline
entireLineMoved & How many full lines in the RCR were moved in the \emph{destination} file? &  Number of lines moved within close proximity can be valuable, for example, a large number of moved lines can indicate a structural rearrangement.\\ \hline
entireLineDeleted & How many full lines in the RCR were deleted from the \emph{source} file? &  Similarly, a large number of deleted lines can indicate a structural change or a larger change.\\ \hline
stringsUpdated & Number of strings in the \emph{source} file that was updated. & Higher number of strings updated in the source file can indicate an `Alternate Output' change. \\ \hline
magicStringsReplaced & Number of Magic strings replaced with variables. & Magic Strings are strings that are specified internally and impact the code behavior. Magic string replacement may indicate control flow-related change. \\ \hline
movedBlocksInIfConditions & Number of blocks within the if conditions were moved in the \emph{destination} file. & Number of moved blocks within the `if' condition indicates control flow-related issues within code. \\ \hline
insertedAssertConditions & Number of asserts inserted in \emph{destination} file & As assert check for condition, insertion of assert condition might indicate functional change. \\ \hline
insertedTryCatch & Number of try-catch nodes inserted in the \emph{destination} file. & Inserted try/catch statement may indicate Validation issues within code. \\ \hline
removedTryCatch & Number of try-catch nodes in the \emph{source} file that was removed in the \emph{destination} file. & Removed try/catch statement may indicate Validation issues or even a Structural change within the code. \\ \hline
updatedValueAssignments & Number of updated statements in the \emph{destination} file where any variables' value was updated. & Variables' value update can indicate logical functional issues. \\ \hline
updatedFunctionArguments & Number of function arguments that were updated in the \emph{destination} file. & Number of updated arguments within a function can provide valuable information about the change. \\ \hline
\multicolumn{3}{l}{\textbf{File-based attributes}}\\ \hline
hasNewFile & A binary variable indicating \emph{destination} file existence. & If 0, then it indicates no change occurred, so the comment might be false positive. \\ \hline
hasOldFile & A binary variable indicating \emph{source} files' existence. & Similarly to the previous one, this variable's value can provide information about a comment. \\ \hline
cyclomaticComplexity & The Cyclomatic Complexity proposed by Mccabe\cite{mccabe1976complexity} of the \emph{source} file. & Cyclomatic complexity can provide vital information regarding the issue of the code or the types of review comments. \\ \hline
commentLOC  & The line number of the \emph{source} file where the review comment was made.  & Comment line number can provide crucial information, such as if the commentLOC is small, then most probably the review comment is targeted for import statements. \\ \hline
\end{tabular}
}
\end{table*}

\subsection{Attribute Selection}
We first consider the code context where the review comment was made. Bosu \emph{et} al. \cite{bosu2015characteristics} found that a code change occurs within a limited proximity (+-10 lines) of code, which they define as the `change trigger'. So, we consider the code context of +-10 lines from the occurrence of a CR comment and define that code segment as Review Comment Range (RCR). Code context can be a crucial feature as CR comments are made to pinpoint the existing code mistake. So, the code context can provide significant insight for classifying review comments. Several recent studies have also focused on utilizing source code context by extracting source code feature vector \cite{alon2019code2vec,feng2020codebert}. We utilize the existing source code vectorization technique and use the code vector as a feature of the machine learning model. The second attribute that we consider is the review comment in the form of natural language. As our goal is to classify CR comments, comment text can be a crucial feature for the classifier. Hence, we also use comment texts as inputs to our models.

\revision{We additionally selected 27 code attributes. Those 27 code attributes were calculated from the \emph{source} and the \emph{destination} files. The \emph{source} file is where the CR comment was made, and the \emph{destination} file is the file that finally merged into the main codebase. If \emph{destination} file is non-existent, no change occurred after the CR comment was made. We selected several Abstract Syntax tree-based (AST)  based attributes, as recent studies have found that AST can better represent source code \cite{zhang2019novel,white2016deep}. Although AST difference-based code attributes were not explored by Fregnan \emph{et} al.~\cite{fregnan2022happens}, those could be crucial to identify the type of recommendation made in a CR comment. For example, if a CR comment suggests no code change, only minor fixes such as updating documentation, there would be no difference between source and destination. If two ASTs have identical structures with different token names, their associated CR comment may be related to refactoring. Moreover, if two ASTs have the same number of tokens and nodes but differ in comparison operators, their CR comment may be related to a logical issue (i.e., functional).
As machine learning features, we have incorporated nine AST-based attributes (e.g., anyDeleted, AnythingInLineMoved, getMovedSrcs).
We also used 14 change-based attributes and four file-based attributes. The change-based and file-based attributes are selected considering our classification goal and the insight we gather from prior studies \cite{fregnan2022happens,fluri2007change,mccabe1976complexity}. Table ~\ref{table:features} presents the 27 selected attributes, a short description for each, and a brief rationale for selection.}

\subsection{Machine Learning Modeling}

\begin{figure}
	\centering  \includegraphics[width=\linewidth]{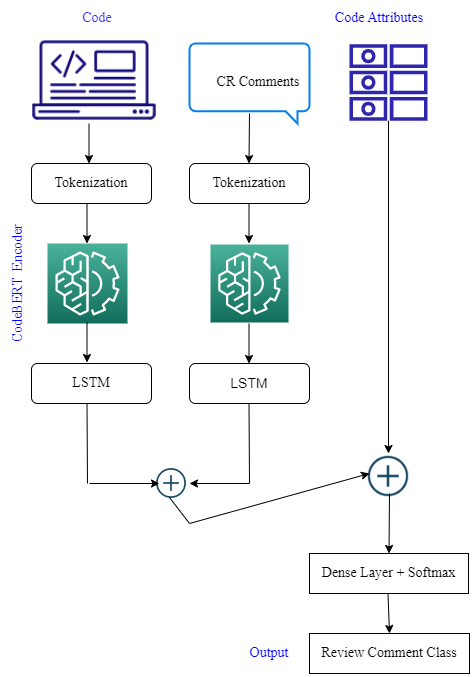}
\caption{Architecture of our proposed Machine Learning model. $\bigoplus$ indicates a concatenation unit in the image. CR Comments $\longrightarrow$ Code Review Comments.}
\label{fig:model_arch}	
%\vspace{-12pt}
\end{figure}

Figure~\ref{fig:model_arch} shows the architecture of our proposed model. The proposed machine-learning model inputs code, CR comments, and code attributes. Text data must be tokenized and converted into vectors to use as inputs for machine learning models. We tokenized the code and CR comments using CodeBERT tokenizer \cite{feng2020codebert} separately. CodeBERT tokenizer provides \textit{input\_ids} and \textit{attention\_mask} for both source code and review comments. For implementing the CodeBERT tokenizer, we used transformers from Hugging Face~\cite{wolf-etal-2020-transformers}. 

For source code and review comment vectorization, we use the CodeBERT encoder\cite{feng2020codebert}. We separately pass the tokenized source code and tokenized review comments to the CodeBERT encoder. CodeBERT is a hybrid pre-trained vectorizer that has been trained with both source code and natural language text. Also, CodeBERT achieved state-of-the-art performance for source code and natural language\cite{feng2020codebert}.
Since we use both source code and CR comments together, CodeBERT may be an excellent option for generating contextualized embedding vectors. After getting the embedding vector for code and CR comments, we pass it through separate LSTM layers. 

 Each embedded source code and review comment vectors are then separately passed through an LSTM \cite{hochreiter1997long} layer. The LSTM layer consists of 50 LSTM cells. The code output and comment output from both the LSTM layers are concatenated together. This concatenated vector is further concatenated with the calculated code attributes of Table \ref{table:features}. The concatenated vector is then passed through a Dense layer with \textit{softmax} activation function. The Dense layer produces the final class that a review comment belongs to. We used the Categorical Cross Entropy \cite{rubinstein2004cross} as the loss function and used Adam optimizer\cite{kingma2014adam} to optimize the loss function. For training the model, we used \emph{batch size} 4 and \emph{epoch} 8. As the code context and review comment generate a 512-dimensional vector, we had to choose a small batch size for training. We also implemented \emph{EarlyStopping} method from the TensorFlow\footnote{https://www.tensorflow.org/} library with 10\% validation data. \emph{EarlyStopping} was performed to avoid overfitting. We performed 10-fold cross-validation, and in each fold, 80\% data were used for training, 10\% were used for validation, and the remaining 10\% were used for testing.

\subsection{Evaluation Metrics}
%Write here
To assess the performance of the proposed model, we compute Precision, Recall, F1-Score, Matthew’s Correlation Coefficient (MCC)\cite{reich1999evaluating}, and overall Model Accuracy.

\subsection{Experiments and Results}
We conducted two types of experiments to evaluate the performance of our proposed approach. First, we experimented on the contribution of code context, CR comment, code features separately, and the contribution while the attributes are combined. We also experimented with different tokenization and embedding on the review comment. The result then leads us to perform an error analysis to understand the model's bias.

\subsubsection{Effect of code context, review comment, and code attributes separately for classifying review comments}

\begin{table*}
\caption{Performance of our proposed tool while code, review comment, and code attributes are considered separately}
\centering \label{table:separate_features}
\resizebox{\linewidth}{!}{
\begin{tabular}{|l|r|r|r|r|r|r|r|r|r|r|r|r|}
\hline
 \multirow{2}{*}{\textbf{Comment class}}  & \multicolumn{4}{c|}{\textbf{Code context only}} & \multicolumn{4}{c|}{\textbf{Review comment only}} & \multicolumn{4}{c|}{\textbf{Code attributes only}} \\ \hhline{~------------}
  & \textbf{Precision} & \textbf{Recall} & \textbf{F1-Score} & \textbf{MCC} & \textbf{Precision} & \textbf{Recall} & \textbf{F1-Score} & \textbf{MCC} & \textbf{Precision} & \textbf{Recall} & \textbf{F1-Score} & \textbf{MCC} \\ \hline
 Discussion & 0.326 & 0.208 & 0.232 & 0.396 & 0.586 & 0.706 & 0.637 & 0.543 & 0.049 & 0.121  & 0.062 & 0.183 \\ \hline
 Documentation & 0.557 & 0.592 & 0.565 & 0.381 &  0.672 &  0.685 & 0.673 & 0.552 & 0.067 & 0.064  & 0.056 & 0.212 \\ \hline
 False Positive & 0.033 & 0.013 & 0.019 & 0.414 & 0.441 & 0.194 & 0.257 & 0.571 & 0.009 & 0.036  & 0.014 & 0.235 \\ \hline
 Functional & 0.176 & 0.055 & 0.056 & 0.414 & 0.416 &  0.327 & 0.337 & 0.568 & 0.068 & 0.297  & 0.093 & 0.224\\ \hline
 Refactoring & 0.415 & 0.719 & 0.523 & 0.283 & 0.602 & 0.624  &  0.604 & 0.520 & 0.255 & 0.452  & 0.263 & 0.145\\ \hline
 \textbf{Model Accuracy} & \multicolumn{4}{r|}{0.418} & \multicolumn{4}{r|}{0.579} & \multicolumn{4}{r|}{0.230} \\ \hline
\end{tabular}
}
%\vspace{-14pt}
\end{table*}
First, we experimented with the impact of each feature separately on the performance of the review comment classification task. Table ~\ref{table:separate_features} presents the result when we experimented with code context, CR comment, and code attributes separately. While Code context was used only, we achieved F1-Score of 0.232, 0.565, 0.019, 0.056, and 0.523 for the \code{Discussion}, \code{Documentation}, \code{False Positive}, \code{Functional}, and \code{Refactoring} classes, respectively. While the Review comment was used only, we achieved F1-Score of 0.637, 0.673, 0.257, 0.337, and 0.604 for the \code{Discussion}, \code{Documentation}, \code{False Positive}, \code{Functional}, and \code{Refactoring} classes respectively. While we used Code attributes only, we achieved F1-Score of 0.062, 0.056, 0.014, 0.093, and 0.263 for the \code{Discussion}, \code{Documentation}, \code{False Positive}, \code{Functional}, and \code{Refactoring} classes respectively. So, we can argue that Code context, Review comment, and Code attributes all contribute to the model performance.

\subsubsection{Effect of different tokenization and vectorization on code review comment}

 CR comment is provided in natural language, so initially we selected pre-trained BERT\cite{kenton2019bert} for review comment tokenization and vectorization task and CodeBERT~\cite{liu2022codebert} for the same for code context. We also experimented using CodeBERT as the $tokenizer +vectorizer$ for review comments as well. 
 The processing of the other two input features, code and code attributes, are kept unchanged, i.e., code context features are always vectorized with CodeBERT. Table~\ref{table:codebert_bert} compares the results between two combinations, i) BERT for   CR comments + CodeBERT for code context, ii) CodeBERT for CR Comments + CodeBERT for code context. Table \ref{table:codebert_bert} suggests that with BERT for CR  + CodeBERT for Code context combination, we achieved an overall model accuracy of 0.502. However, with CodeBERT for CR comments + CodeBET for code context combination, the model achieves an accuracy of 0.593, which is almost 10 percent higher than the other combination. 

\begin{table*}
\caption{Performance comparison while BERT and CodeBERT are used separately for review comment tokenization and vectorization task}
\centering \label{table:codebert_bert}
\begin{tabular}{|l|r|r|r|r|r|r|r|r|}
\hline
  \multirow{3}{*}{\textbf{Comment class}}  & \multicolumn{4}{c|}{BERT} & \multicolumn{4}{c|}{CodeBERT} \\ 
   & \multicolumn{4}{c|}{$(tokenizer+vectorizer)$} & \multicolumn{4}{c|}{$(tokenizer+vectorizer)$} \\ \cline{2-9}
  & \textbf{Precision} & \textbf{Recall} & \textbf{F1-Score} & \textbf{MCC} & \textbf{Precision} & \textbf{Recall} & \textbf{F1-Score} & \textbf{MCC} \\ \hline
 Discussion & 0.516 & 0.570 &  0.522 & 0.449 & \textbf{0.611}  & \textbf{0.685}  & \textbf{0.638} & \textbf{0.547} \\ \hline
 Documentation & 0.670 & 0.634 &  0.644 & 0.463 & \textbf{0.704}  &  \textbf{0.762} & \textbf{0.723} & \textbf{0.556} \\ \hline
 False Positive & 0.124 &  0.041 & 0.056 & 0.487 & \textbf{0.376} & \textbf{0.158}  & \textbf{0.207} & \textbf{0.576} \\ \hline
 Functional & 0.380 &  0.116 & 0.162 & 0.483 & \textbf{0.408}  & \textbf{0.358}  & \textbf{0.365} & \textbf{0.572} \\ \hline
 Refactoring &  0.461 & 0.632 & 0.523 & 0.409 & \textbf{0.628}  &  \textbf{0.633} & \textbf{0.616} & \textbf{0.527} \\ \hline
 \textbf{Model Accuracy} & \multicolumn{4}{r|}{0.502} & \multicolumn{4}{r|}{\textbf{0.593}} \\ \hline
\end{tabular}
%\vspace{-14pt}
\end{table*}

\begin{figure}
	\centering  \includegraphics[width=9 cm]{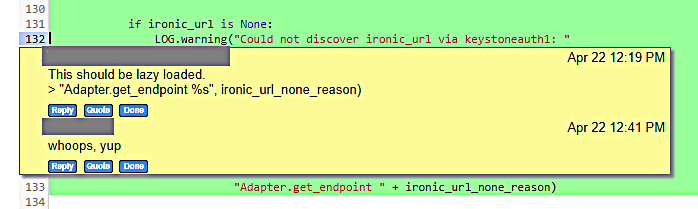}
\caption{Example of a review comment where the reviewer presents a code line to pinpoint the issue better. This example has been taken from the OpenDev Nova project.}
\label{fig:review_with_code}	
\end{figure}

From Table \ref{table:codebert_bert}, \revision{the result suggests} that CodeBERT performs much better than the BERT for review comment tokenization and vectorization task. We were initially surprised by the findings and decided to \revision{manually inspect the CR comments. We selected 200 review comments randomly from our dataset for manual inspection.} We found that reviewers present code chunks in many CR comments for pinpointing the issue to the code author. Figure~\ref{fig:review_with_code} presents an example from the OpenDev Nova project. CR comments may contain natural language as well as programming language segments. Similarly,  CodeBERT, is trained on both natural language and source code. Therefore, CodeBERT performs better than BERT for review comment tokenization and vectorization tasks. 

\subsubsection{Error Analysis}

\begin{table*}
\caption{Confusion Matrix for error analysis}
\centering \label{table:confusion_matrix}
%\resizebox{\linewidth}{!}{
\begin{tabular}{|l|r|r|r|r|r|}
\hline
 \backslashbox{Ground truth}{Predicted} & \textbf{Discussion } & \textbf{Documentation} & \textbf{False Positive} & \textbf{Functional} & \textbf{Refactoring} \\ \hline
\textbf{Discussion} & 306 & 29 & 18 & 28 & 64 \\ \hline
\textbf{Documentation} & 32 & 292 & 3 & 8 & 52 \\ \hline
\textbf{False Positive} & 43 & 29 & 24 & 26 & 36 \\ \hline
\textbf{Functional} & 37 & 15 & 13 & 85 & 90 \\ \hline
\textbf{Refactoring} & 94 & 57 & 13 & 57 & 377 \\ \hline

\end{tabular}
%}
%\vspace{-14pt}
\end{table*}

We analyzed the misclassifications qualitatively and quantitatively to better understand our model's performances. Table~\ref{table:confusion_matrix} shows the confusion matrix from 10-fold cross-validation for the best model. The confusion matrix shows how many samples of each class are correctly predicted and how many samples are classified into other classes. Since we took the classified instances for the confusion matrix from 10 folds, we combined all the samples for the test set and put their sums on the confusion matrix. The confusion matrix of Table~\ref{table:confusion_matrix} also provides us insights into the model's biases. 

\noindent \textbf{False Positive:} According to the confusion matrix, out of 158 \code{False Positive} samples, 24 cases are accurately identified, whereas 43 samples are misclassified as \code{Discussion}, 29 samples are misclassified as \code{Documentation}, 36 samples are detected as \code{Refactoring}, and 26 samples as \code{Functional} class. So, our classifier is biased toward the other four classes in a similar ratio for the' \code{False Positive} comments. We manually analyzed those cases, and an example of a \code{False Positive} review comment that is classified as a \code{Functional}, ``\textit{A simple unit test would assert that we don't call service\_get\_all if enabled\_services is passed in.}" This review comment addresses a Functional comment, but the concern was not valid for that circumstance. Therefore, our labelers marked it as a \code{False Positive}. 

\noindent \textbf{Discussion:} We observe biases towards the \code{Refactoring} class while predicting the \code{Discussion} class comments. For example, \revision{``\textit{We should make this a public method now yeah?}"} belongs to the \code{Discussion} class but our model has misclassified it as a \code{Refactoring}. 

\noindent \revision{\textbf{Documentation:}  We found biases of the \code{Documentation} class towards the \code{Refactoring} class. For example, ``\textit{nit: add a TODO to remove this proxy and just have callers hit the necessary clients directly.}'' is a \code{Documentation} comment that our classifier misclassified as a \code{Refactoring}.}

\noindent \textbf{Refactoring:} From our evaluation, we found that many \code{Refactoring} samples are misclassified as a \code{ Discussions}. For instance, \revision{``\textit{Is anything else using this method now or can we remove it also?}"} is misidentified in the \code{Discussion} class category. 

\noindent \textbf{Functional:} In cases, where \code{Functional} samples were misclassified, they were more likely to be predicted as the \code{Refactoring} category. An example where a \code{Functional} comment is misclassified as a \code{Refactoring} comment, \revision{``\textit{I would lower this too to be safe, but I guess it's not going to change.}"}

\begin{boxedtext}
\textbf{Key takeaway 1:} \emph{The proposed machine learning model can classify CR comments with 59.3\% accuracy.  The results of our evaluation also suggest that all  three categories of input features (i.e., code attributes, code context, and comment text) contribute to improving model performance.}
\end{boxedtext}

\begin{boxedtext}
\textbf{Key takeaway 2:} \emph{Although review comments are provided in natural language, they often contain programming language lines and keywords. So, for the tokenization and vectorization of CR comments, we recommend using a hybrid model that is trained both in natural language and programming language, such as CodeBERT\cite{feng2020codebert}.}
\end{boxedtext}

\vspace{5pt}
\section{(RQ2) Comparison against current state-of-the-art for review comment classification task}
\label{sec:rq2}

This section presents the replication of Fregnan et al.\cite{fregnan2022happens} work and the performance comparison between the two approaches.

\subsection{Replication}
The work of Fregnan et al.\cite{fregnan2022happens} proposed models to classify review-induced changes into four categories. Although most CR comment categories trigger changes, some categories, such as false positive, discussion, or praise, do not. Even though our classifier has a slightly different goal, we can train a model using their approach and our dataset for CR comment classification. Therefore, we retrained Fregnan et al.'s\cite{fregnan2022happens} model to compare the performance of our approach against theirs. We followed their approach as closely as possible for replication.  Despite the attributes of Fregnan \emph{et} al.\cite{fregnan2022happens} being calculated for Java code,  and our selected project being written in Python, we were able to calculate the same set of attributes 

\revision{ Since Fregnan \emph{et} al.\cite{fregnan2022happens} provide adequate descriptions of the selected attributes and have made their replication package publicly available~\cite{enrico_fregnan_2021_5592254}, we were able to write a script to extract their selected attributes for our dataset. We use our five-class category label for each CR comment as the dependent variable, which we computed in the data preparation step. Thus, we could utilize their approach in our context and obtain results as a five-class classification.} As the Random Forest algorithm performed the best on classifying review-induced changes during their evaluation, we used this algorithm with 10-fold cross-validation for comparison.

\begin{table*}
\caption{Comparison between our proposed approach with the approach of Fregnan \emph{et} al.\cite{fregnan2022happens}}
\centering \label{table:comparison_fregnan}
\begin{tabular}{|l|r|r|r|r|r|r|r|r|}
\hline
 \multirow{2}{*}{\textbf{Comment class}}   & \multicolumn{4}{c|}{\textbf{Fregnan \emph{et} al.'s Approach}} & \multicolumn{4}{c|}{\textbf{Proposed Approach}} \\ \hhline{~--------}
 & \textbf{Precision} & \textbf{Recall} & \textbf{F1-Score} & \textbf{MCC} & \textbf{Precision} & \textbf{Recall} & \textbf{F1-Score} & \textbf{MCC}  \\ \hline
 Discussion & 0.355 & 0.373 &  0.362 & 0.368 & \textbf{0.611}  & \textbf{0.685}  & \textbf{0.638} & \textbf{0.547} \\ \hline
 Documentation & 0.477 & 0.497 &  0.483 & 0.382 & \textbf{0.704}  &  \textbf{0.762} & \textbf{0.723} & \textbf{0.556} \\ \hline
 False Positive & 0.283 &  0.040 & 0.068 & 0.410 & \textbf{0.376} & \textbf{0.158}  & \textbf{0.207} & \textbf{0.576} \\ \hline
 Functional & 0.297 &  0.093 & 0.139 & 0.408 & \textbf{0.408}  & \textbf{0.358}  & \textbf{0.365} & \textbf{0.572} \\ \hline
 Refactoring &  0.415 & 0.592 & 0.486 & 0.307 & \textbf{0.628}  &  \textbf{0.633} & \textbf{0.616} & \textbf{0.527} \\ \hline
 \textbf{Model Accuracy} & \multicolumn{4}{r|}{0.406} & \multicolumn{4}{r|}{\textbf{0.593}} \\ \hline
\end{tabular}
%\vspace{-14pt}
\end{table*}

\subsection{Result Comparison}
Table~\ref{table:comparison_fregnan} presents the result of the Fregnan \emph{et} al.\cite{fregnan2022happens} approach and our proposed approach. The values presented in Bold font represent the best value between the two models. The results from Table \ref{table:comparison_fregnan} indicate that for all the measures, our proposed model performs better than the approach of Fregnan \emph{et} al.\cite{fregnan2022happens}. 

From Table \ref{table:comparison_fregnan}, we can see that, for the \code{Discussion}, \code{Documentation}, \code{False Positive}, \code{Functional}, and \code{Refactoring} classes, Fregnan \emph{et} al.\cite{fregnan2022happens} approach achieved an F1-Score of `0.362', `0.483', `0.068', `0.139', and `0.486' respectively. Whereas, for the \code{Discussion}, \code{Documentation}, \code{False Positive}, \code{Functional}, and \code{Refactoring} classes, we have achieved an F1-Score of `0.638', `0.723', `0.207', `0.365', and `0.616' respectively. Fregnan \emph{et} al.\cite{fregnan2022happens} approach achieved an overall model accuracy of 0.406, whereas we achieved an overall model accuracy of 0.593. Our proposed approach has achieved a significant performance improvement for all five comment classes and overall model accuracy.
%Only for the `Evolvability' class, Fregnan et al.\cite{fregnan2022happens} approach has achieved a higher recall value than our model. 

\begin{boxedtext}
\textbf{Key takeaway 3:} \emph{Our proposed approach acheives better performance than Fregnan \emph{et} al.\cite{fregnan2022happens} for the review comment classification task across all measures. Fregnan \emph{et} al.\cite{fregnan2022happens} approach shows lower performance for the \code{False Positive} class. Although our proposed approach achieves a better performance than Fregnan et al.\cite{fregnan2022happens} for the \code{False Positive} class; still, the performance needs significant improvements. Further study is required for better analyzing and improving the detection of \code{False Positive} review comments.}
\end{boxedtext}

\vspace{5pt}
\section{Implications}
\label{sec:discussion}
This section presents the insights obtained from this study, potential future research directions, and recommendations for CR practitioners.

\vspace{2pt}
\noindent \emph{1. Improving classification performance:}
Our proposed model result shows that the model performs better for the \code{Discussion}, \revision{\code{Documentation}, \code{Refactoring},} and \code{Functional} classes than for the \code{False Positive} class. The approach of Fregnan et al.\cite{fregnan2022happens} shows an even worse performance for the \code{False Positive} class than our model. This result leads us to further investigate the reasons behind such poor showings for the \code{False Positive} class. We found that only 8.64\% comments in our dataset belong to the \code{False Positive} class. 
This underrepresentation of  the \code{False positive} in our sample may be a possible cause. While we did not explore class-balancing techniques such as oversampling, that may be a possible direction to improve performances.
After examining false positive comments closely, we found that sometimes review comments pointing to issues belonging to other classes can become \code{False Positive} if refuted. For example, a reviewer has alluded to a \code{Functional} issue, but the concern is not valid for the current scenario. Therefore, while the same comment may have been \code{Functional} for another context, it is a \code{False positive} under the current scenario. This observation also shows the need to consider code context characteristics to classify CR comments. Moreover, the results of our error analysis also reveal cross miss-categorizations among the five classes. This analysis can also help identify additional features to distinguish between such pairs.

\vspace{2pt}
\noindent \emph{2. Selection of pre-trained vectorizers:}
From our experiment with two different combinations of $tokenizer+vectorizer$, we found that tokenization and vectorization can affect the performance of the review comment classifier. Code review comments contain both natural language and programming language tokens. So, a multipurpose model, such as CodeBERT~\cite{liu2022codebert} that is trained both in natural language and programming language performs better for review comment tokenization and classification task. Therefore, we  recommend evaluating multipurpose pre-trained models in conjunction with general-purpose models for SE domain-specific NLP pipelines.
\vspace{2pt}
\noindent \emph{3. Recommendations for Practitioners: }
To improve CR effectiveness, an organization needs to define a set of metrics and measure those to identify potential improvement areas~\cite{hasan2021using}. Projects can leverage our model to track CR performances, such as counting the number of issues belonging to various categories, issue category-wise contributions from each reviewer for a project, and the ratios of defects escaping. These insights can help managers to identify potential improvement areas and adopt new initiatives~\cite{hasan2021using}.
Besides analytic supports, our model can also help authors prioritize CR comments. For example, a CR may receive ten comments from a reviewer, where eight are minor nitpicking issues, and the two remaining ones are critical defects. Using the code context and CR comment text, our model can automatically identify functional issues and help authors prioritize those accordingly. Authors can also leverage an analytic platform built on our model to identify reviewers who have more frequently identified the type of issues (e.g., \code{Refactoring} vs. \code{Functional}) that they are seeking for the current CR

\vspace{2pt}
\noindent \emph{4. Recommendations for researchers: }
Although our model improves the current state-of-the-art (SOTA) for CR comment classification by almost 20\% in terms of accuracy,  our best model has only 59.3\% accuracy. While a five-category automated classification is more difficult than binary classifications, we believe there are significant improvement opportunities in this direction. Potential directions for improvements include training on larger-scale datasets, including balanced ratios of issues representing various categories and evaluating with oversampling techniques. 

In addition, to improving this classifier, we believe our model can help build better review automation tools. As most of the CR comments belong to trivial issues, CR automation tools~\cite{thongtanunam2022autotransform,tufano2021towards,li2022codereviewer,tufano2022using,tufano2019learning} trained on randomly curated datasets are more likely to amplify such issues and miss the rare yet more crucial ones.
Future CR automation tools can leverage our model to build a more balanced sample, as manual labeling is time-consuming. 
Finally, existing history-based automated reviewer recommendation systems~\cite{zanjani2015automatically,thongtanunam2015should,rahman2016correct,pandya2022corms} consider all prior review interactions equally, which may not be the best approach. Our automated model can analyze reviewers' prior feedback history and provide higher priority to reviewers who previously provided more useful (e.g., \code{Functional}) feedback.

%\vspace{2pt}
%\noindent \emph{5. Lessons Learned from the Study: }
%After conducting different experiments, we found only the code review comment itself can provide enough context to classify the comment. As we prioritized F1-score and model accuracy, from our point of view, the ensemble of code context, review comment, and code attributes is the best-performing model. But running our proposed model requires a lot of time and computational resources. If the computational resource is an issue, then only the review comment processing channel (review comment $\rightarrow$ CodeBERT tokenizer $\rightarrow$ CodeBERT vectorizer $\rightarrow$ LSTM layer $\rightarrow$ Dense Layer + Softmax) of our model architecture can be utilized for classifying review comments. 

\section{Threats to Validity}
\label{sec:thrests}

The following subsections detail potential threats to the validity of this study and our countermeasures to mitigate those threats.

\vspace{2pt}
\noindent \emph{A. Internal Validity:}
To compare the performance of our approach with Fregnan \emph{et} al.\cite{fregnan2022happens}, we retrained their model on our dataset. A possible threat might appear during this retraining. To avoid any bias, we followed their approach as described in their paper. Although their code is publicly available, we could not directly use it because their attributes were calculated for the Java project, while our subject (i.e., OpenDev Nova) is written in Python. However, this threat may be minimal, as we used standard libraries to compute the attributes and followed their definitions. We also performed 10-fold cross-validation to mitigate biases introduced by using fixed training and testing samples.

\vspace{2pt}
\noindent \emph{B. Construct Validity:}
A significant threat to construct validity is the manual categorization process. For categorizing the code review comments, we performed a manual categorization. Using a rubric adopted from prior studies, we categorized each CR comment into one of the 17 categories. To mitigate the effect of human error, two authors did the categorization independently. We measured the agreement between the two annotators using Cohen's kappa\cite{cohen1960coefficient} value. The manual labeling task achieved a kappa value of 0.68, which shows a `Substantial' agreement. A third author then resolved the conflicts between the two authors' labeling. Therefore, this threat to validity remains minimal.
Another potential threat to the construct validity is the features that were selected for the machine learning approach. Sometimes, used features might affect the performance of the model negatively. To mitigate the effect, we experimented separately with each type of feature we used in machine learning modeling. The result validates that each feature category contributes to the model's performance. Also, to mitigate the effect of improper tokenization and vectorization of review comments, we experimented with both BERT and CodeBERT. 

\vspace{2pt}
\noindent \emph{C. External Validity:}
The main threat to external validity is the generalizability of the approach. Our approach requires manual labeling, which is extremely time-consuming. As a result, we could not incorporate multiple projects into this experiment.
Even so, as characteristics vary from project to project, one cannot claim the generalizability of an approach with multiple projects. 
Regardless, this study has proposed an approach that can be utilized flexibly to develop project-specific models. For example, if, for some projects, the code context does not contribute to the overall performance, then the processing channel of code context can be discarded, and the remaining portion of our architecture can be utilized for CR comment classification tasks. To facilitate the replication of our work, we have made the dataset and the code publicly available.

\vspace{2pt}
\noindent \emph{D. Conclusion Validity}
First, selecting metrics to evaluate a model's performance can also threaten conclusion validity. We have used five different, widely used performance measuring metrics to mitigate this threat: precision, recall, F1-score, MCC, and model accuracy. Therefore, we do not anticipate any biases arising from our metrics. Second, machine learning models can show biased results when a fixed portion of data is selected for testing and training. To mitigate the effect of such bias, we performed a 10-fold cross-validation. Ten-fold cross-validation ensured that every portion of the dataset was utilized separately for the training and testing process. 
Finally, overfitting can be a significant threat to machine learning modeling. Overfitting occurs when a machine learning model performs well on training data but poorly on test data. We implemented the \emph{EarlyStopping} method to avoid overfitting with 10\%  validation data. This method monitors the model performance on validation data and stops the training process once the convergence of the training process halts.

\section{Conclusion}
\label{sec:conclusion}

We have proposed a machine-learning approach to classify code review comments in this work. We collected 1,828 code review comments from the OpenDev Nova project and manually classified the code review comments using the prior review comment classification scheme. We experimented with different features that might be used for the review comment classification task and reported the contribution of those features separately. We then combined those features and developed a machine-learning model that uses those features for classifying code review comments. 

Our proposed approach takes code context, code review comments, and a set of code attributes for classifying code review comments. We achieved an overall model accuracy of 59.3\%. Experimental results suggest our proposed approach can classify code review comments better than the existing change classification approach proposed by Fregnan \emph{et} al.\cite{fregnan2022happens}. We also experimented with how different tokenization and vectorization approaches can influence the performance of a review comment classifier. Finally, we presented the insight gathered by conducting this study, opportunities for researchers, and draw recommendations for practitioners.
\section*{Acknowledgement}

Work conducted for this research is partially supported by the US National Science Foundation under Grant No. 1850475. Any opinions, findings, conclusions, or recommendations expressed in this material are those of the author(s) and do not necessarily reflect the views of the National Science Foundation. We thank Md. Toufikuzzaman for his assistance during the data mining process.

\bibliographystyle{IEEETran}
\balance
\bibliography{bibliography}

\end{document}